\documentclass[prd,aps,showpacs,preprintnumbers,amssymb]{revtex4}
\usepackage{axodraw}
\usepackage{color}
\usepackage{epsf}

\def\e3p{$\eta \rightarrow 3 \pi$}

\begin{document}
\title{%
\hfill{\normalsize\vbox{%
\hbox{}
 }}\\
{A purely quark Lagrangian from QCD}}

\author{Amir H. Fariborz
$^{\it \bf a}$~\footnote[1]{Email:
 fariboa@sunyit.edu}}

\author{Renata Jora
$^{\it \bf b}$~\footnote[2]{Email:
 rjora@theory.nipne.ro}}

\affiliation{$^{\bf \it a}$ Department of Matemathics/Physics, SUNY Polytechnic Institute, Utica, NY 13502, USA}
\affiliation{$^{\bf \it b}$ National Institute of Physics and Nuclear Engineering PO Box MG-6, Bucharest-Magurele, Romania}

\date{\today}

\begin{abstract}
We present a method for determining  a purely quark Lagrangian by mocking up the QCD partition function for large gauge couplings $g$. The resulting effective theory displays all the symmetries of low energy QCD and can be potentially used to explore hadron properties.
\end{abstract}
\pacs{12.38.Aw, 12.38.Lg, 12.39.Fe}
\maketitle

\section{Introduction}

Quantum chromodynamics is an $SU(3)$ gauge theory which contains quarks and gluons that interact with each other. Whereas at high energies the coupling constant is small and one can adequately describe the reality by expanding perturbatively in the coupling constant it is still not completely clear  how one can depict the low-energy regime where the coupling constant is large and quarks form bound states of mesons and baryons. Various methods and Lagrangians have been proposed to describe the low energy dynamics: While some of them like the Nambu Jona Lasinio model \cite{NJL1}, \cite{NJL2} are still based on fermions as fundamental degrees of freedom, most of them consider the mesons and baryons as the starting point. In the latter case one then constructs effective low-energy models which are widely based on the symmetries of the QCD Lagrangian like the $SU(3)_L\times SU(3)_R$ chiral symmetry, $U(1)_V $ or $U(1)_A$ (as well as assumptions  about QCD vacuum and potential) embedded in the quark flavor sector of QCD. Models of low energy QCD with notable results regarding hadrons properties and interactions include chiral perturbation theory \cite{Leutwyler1},\cite{Leutwyler2}, linear and nonlinear sigma models \cite{SU}-\cite{FJS08} or other symmetry induced effective Lagrangians \cite{mixing}-\cite{tsm}. Thus one can conclude that chiral symmetry whether spontaneously or explicitly broken is the major ingredient for building a low energy QCD effective theory.

In the present work however we shall consider a completely different point of view: we will not rely on symmetries but instead we shall build a purely dynamical Lagrangian  and obtain the chiral symmetry as a derivative of the method.  Our main tool is the QCD partition function from which one can derive all the properties of the particles and interactions.  We start by considering an alternative description of the partition function after integrating out the quark degrees of freedom.  As a consequence the original quarks will be replaced by copies (that describe the constituent quarks) with the same masses and quantum numbers but with different Lagrangian and thus interactions. In the end we shall obtain a Lagrangian where the gluon degrees of freedom have been eliminated and that contains only the quarks and their subsequent interactions. Note that our method implies only manipulation and calculations of the partition function and does not involve any loop computation.

We consider the effective fermion Lagrangian obtained in this paper as a purely theoretical one and as  a first step towards a more comprehensive approach. The connection with the phenomenology of the bound states of mesons and baryons will be performed and finalized in a future work.

\section{the QCD Lagrangian and partition function}

We start with the gauge fixed QCD Lagrangian with $N$ colors and $N_f$ flavors in the fundamental representation:
\begin{eqnarray}
{\cal L}=-\frac{1}{4}(F^a_{\mu\nu})^2+\bar{c}^a(-\partial^{\mu}\partial_{\mu}-g f^{abc}\partial^{\mu}A^b_{\mu})c^c+\sum_f\bar{\Psi}_f(i\gamma^{\mu}D_{\mu}-m_f)\Psi_f,
\label{rez5467}
\end{eqnarray}
where,
\begin{eqnarray}
F^a_{\mu\nu}=\partial_{\mu}A^a_{\nu}-\partial_{\nu}A^a_{\mu}+gf^{abc}A^b_{\mu}A^c_{\nu},
\label{ten45678}
\end{eqnarray}
and,
\begin{eqnarray}
D_{\mu}=\partial_{\mu}-igA^a_{\mu}t^a.
\label{cov554}
\end{eqnarray}
Here as usual $t^a$ are the generators of the group $SU(N)$ in the fundamental representation.

We shall ignore the ghosts in what follows as they do not contribute essentially to our arguments.
First we  separate the Lagrangian in two pieces:
\begin{eqnarray}
&&{\cal L}_1=\sum_f\bar{\Psi}_f(i\gamma^{\mu}\partial_{\mu}-m_f+g\gamma^{\mu}t^aA^a_{\mu})\Psi_f
\nonumber\\
&&{\cal L}_2=-\frac{1}{4}(F^a_{\mu\nu})^2.
\label{sep998}
\end{eqnarray}

The QCD partition function has the form:
\begin{eqnarray}
Z_0=\int  d A^a_{\mu}(x) d \bar{c}^b(x)d c^d(x)d\bar{\Psi}_{fl}{x} d\Psi_{fl}(x)\exp[i\int d^4x {\cal L}],
\label{part456}
\end{eqnarray}
where the usual rules of the path integration apply.  Here $f$ is the flavor index whereas $l$ is the color one. Then one can integrate over the fermion variables to obtain:

\begin{eqnarray}
Z_0=\int d A^a_{\mu}(x)\prod_f\det\left[i\gamma^{\mu}\partial_{\mu}-m_f+g\gamma^{\mu}t^aA^a_{\mu}\right]\exp[\int d^4x {\cal L}_2].
\label{part456}
\end{eqnarray}

\section{A simplified approach}

We shall discuss in particular the real life case of $N=3$, $N_f=3$ corresponding to QCD with three light flavors. We introduce the fermionic current that couples with the gluon field for each fermion species:
\begin{eqnarray}
J_{\mu f }^a=\bar{\Psi}_f\gamma^{\mu}t^a\Psi_f.
\label{current66}
\end{eqnarray}
For one generation of fermions there are $8\times3$ degrees of freedom for each fermion where $8$ represents space time degrees of freedom of an off-shell fermion and $3$ the number of colors. Overall we have $24N_f$ degrees of freedom for all fermions.

We consider a single flavor and next extend our arguments to $N_f$ flavors. We start by making a change of variables from the elementary fermion degrees of freedom to the composite current $J^a_{\mu}$. We first mention that in the case of anticommuting variables the Jacobian appear with an inverse power as compared to the case of commuting ones.  Note that for this change of variables  to make sense we need $24$ degrees of freedom for $J^a_{\mu}$ instead of the $32$ that one might obtain from a simple counting. This means that we need something similar to a gauge conditon which we choose to be  $\partial_{\mu}J^a_{\mu}=\omega$ where $\omega$ is an arbitrary function as in the more standard case of a gauge field. Since there are exactly $8$ constraints one obtains the desired matching of $24$ degrees of freedom for both the fermion and vector boson variables. Then the change of variable is:
\begin{eqnarray}
\int d\bar{\Psi} d \Psi \rightarrow \int d J^a_{\mu}\delta\left(\partial^{\mu}J^a_{\mu}-\omega\right) \left|\frac{d J^a_{\mu}}{d \Psi_{i}}\right|
\label{rez554}
\end{eqnarray}
 The Jacobian in Eq. (\ref{rez554}) is a determinant with dimension $24$. Since any arbitrary derivative $\frac{d J^a_{\mu}}{d \Psi_{in}}=\bar{\Psi}_{jm}(\gamma_{\mu})_{ji}(t^a)_{mn}$ (here the first indices are space time whereas the second ones are color) contains a fermion variable  then the determinant will contain products of $24$ fermion variables. Taking into account that there are exactly $24$ distinct fermion variables and the  anticommuting nature of these we conclude that the actual determinant will be given exactly by the product of the $24$ distinct variables (since those term that contain a repeated variable will be zero) times an irrelevant constant factor. Thus:
\begin{eqnarray}
\left|\frac{d J^a_{\mu}}{d \Psi_{i}}\right|={\rm const} \prod_{i,m}\Psi_{im}.
\label{rez44232}
\end{eqnarray}
Then one can also write quite safely:
\begin{eqnarray}
\left|\frac{d J^a_{\mu}}{d \Psi_{i}}\right|\rightarrow \det[t^a\gamma^{\mu}J^a_{\mu}]
\label{re4443}
\end{eqnarray}
Here the determinant is taken in the space $\gamma^{\mu}t^a$ so it has the dimension $12$ which leads exactly to a product of $24$ distinct fermion variables. Note that at each point we take into account the composite nature of $J^a_{\mu}$.

We need to extend our arguments to $N_f=3$ flavors.  By applying the previous approach (or simply consider a product over the umber of flavors we obtain:
\begin{eqnarray}
\int \prod_f d \bar{\Psi}_f d \Psi_f \rightarrow \int \prod_fd J^a_{\mu f} \det[\gamma^{\mu}t^a J^a_{\mu f}] \delta\left(\partial^{\mu}J^a_{\mu f}-\omega\right).
\label{rez442332}
\end{eqnarray}
It is more convenient however to  write:
\begin{eqnarray}
\prod_f \det[\gamma^{\mu}t^aJ^a_{\mu f}]\rightarrow{\rm const}\times \det[t^a\gamma^{\mu}\sum_fJ^a_{\mu f}]^3
\label{res53453}
\end{eqnarray}
The above relation is due to the fact that the above determinant contains a  product of $72$ fermion variables. But this is the total number of fermion variables so that each term that contains a variable twice is zero. Thus the determinant will be a constant times the product of $72$ distinct fermion variables. But this is also what one would obtain from Eq. (\ref{rez442332}) so Eq. (\ref{res53453}) is correct.

This being settled we need to write the full partition function in Eq. (\ref{part456}) in terms of the new variables $J^a_{\mu f}$. As we mentioned previously  the change of variables makes sense only with the additional constraint on the field $J^a_{\mu}$. We shall consider a particular case of it with $\omega(x)=0$.  On the other side this constraint is equivalent to constraining the free equation of motion of the fermion field. Thus it can be fulfilled only if,
\begin{eqnarray}
i\gamma^{\mu}\partial_{\mu}\Psi-m\Psi=\frac{a(x)^2}{M}\Psi(x).
\label{eq4423}
\end{eqnarray}
We use the function $\frac{a^2(x)}{M}$ (where $M$ is an arbitrary scale) instead of simply $a(x)$ because we do not want to further constrain the fermion fields $\Psi$.
Then the kinetic term for the fermion field will be replaced in the lagrangian by:
\begin{eqnarray}
\bar{\Psi}(i\gamma^{\mu}\partial_{\mu}-m)\Psi=\frac{a^2(x)}{M}\bar{\Psi}\Psi.
\label{kin65775}
\end{eqnarray}

Furthermore since the function $a(x)$ is arbitrary one can use the equation of motion to eliminate the kinetic term altogether from the Lagrangian.

The partition function in Eq. (\ref{part456}) will thus become:
\begin{eqnarray}
Z_1=\int d A^a_{\mu}\prod_f d J^b_{\rho f} \det[t^a\gamma^{\mu}\sum_fJ^a_{\mu f}]^3\exp\left[i\int d^4 x[ g\sum_fJ^a_{\mu f}A^{a\mu}+{\cal L}_2]\right]
\label{rez5546789}
\end{eqnarray}

The same partition function is obtained if one introduces three flavors of fermion  copies $\chi_f$, $\bar{\chi}_f$ of the original quarks such that:
\begin{eqnarray}
&&Z_1=\int d A^a_{\mu} \prod_{f'} d J^b_{\rho f'}\prod_f d \bar{\chi}_f d\chi_f\exp\left[-i\int d^4 x \frac{1}{M^2}\sum_f\bar{\chi}_f \gamma^{\mu}t^a[\sum_{f'}J^a_{\mu f'}]\chi_f\right]\times
\nonumber\\
&&\exp\left[i\int d^4 x[ \sum_{f'}J^a_{\mu f'}gA^{a\mu}+{\cal L}_2]\right].
\label{ret5665}
\end{eqnarray}

We make the change of variable $K^a_{\mu 1}=\sum_f J^a_{\mu f}$, $K^a_{\mu 2}=J^a_{\mu 2}$, $K^a_{\mu 3}=J^a_{\mu 3}$ to determine:
\begin{eqnarray}
&&Z_1=\int d A^d_{\sigma} d K^a_{\mu 1} K^b_{\rho 2} d K^c_{\nu 3}\prod_f d \bar{\chi}_f d\chi_f\exp\left[i\int d^4 x \sum_f\frac{1}{M^2}\bar{\chi}_f \gamma^{\mu}t^aK^a_{\mu 1}\chi_f\right]\times
\exp\left[i\int d^4 x[ gK^a_{\mu 1}A^{a\mu}+{\cal L}_2]\right]=
\nonumber\\
&&\approx \int d A^a_{\mu} d K^b_{\rho 2} d K^c_{\nu 3}\delta\left(A^{a\mu}+\frac{1}{gM^2}\sum_f\bar{\chi}_f\gamma^{\mu}t^a\chi_f\right)\exp[i\int d^4x {\cal L}_2],
\label{rez226645}
\end{eqnarray}
where the integral over $K^a_{\mu 1}$ is a delta function.
Here one can drop the unwanted integrals over $K^b_{\rho 2}$ and $K^c_{\nu 3}$ since they do not contribute in  any process and apply the delta function to obtain that the effective Lagrangian is just:
\begin{eqnarray}
{\cal L}_{eff}={\cal L}_2\left(A^{a\mu}=-\frac{1}{gM^2}\sum_f\bar{\chi}_f\gamma^{\mu}t^a\chi_f\right).
\label{rez5546}
\end{eqnarray}

Although this procedure seems oversimplifying it gives a correct glimpse of what kind of Lagrangian we should expect in terms of the current $J^a_{\mu}$ if the fermion kinetic term is neglected.   The next section will contain a more general and comprehensive approach.

\section{A comprehensive approach}

We  start with Eq. (\ref{part456}) which we rewrite  here for completeness:
\begin{eqnarray}
Z_0=\int d A^a_{\mu}(x)\prod_f\det\left[i\gamma^{\mu}\partial_{\mu}-m_f+g\gamma^{\mu}t^aA^a_{\mu}\right]\exp\left[i\int d^4x {\cal L}_2\right].
\label{part7456}
\end{eqnarray}
We  shall now try to reproduce the above partition function using a different set of variables.

We first need an identity that we shall prove in what follows. Consider the integral:
\begin{eqnarray}
Z_x= \int d \bar{x}_i d x_i d \bar{y}_j d y _j d J_k d S_m\exp[i[\frac{1}{M^5}\bar{x}A^kxJ^k+\frac{1}{M^4}\bar{y}B^kyS^ky+\frac{1}{M^2}\sum_{k}S^kJ^k]].
\label{intyytu6}
\end{eqnarray}

Here $x_i$, $y_i$ and their conjugates are each a set of $n$ Grassmann variables which we shall assume have the mass dimension $m^{3/2}$ and $J_k$, $S_k$ bare a set of regular commuting ones each with mass dimension $m$. The index $k$ goes also from $1$ to $m$ and the matrices $A^k$, $B^k$ are each a set of $m$  $(n\times n)$ matrices where $A^k$ have mass dimension $m$ and $B^k$ have mass dimension $m^0$.  We will solve first the integral over the fermion fields to get:
\begin{eqnarray}
&&Z_x=\int  dJ^k dS^k \det[i\frac{1}{M^5}A^kJ^k]\det[i\frac{1}{M^4}B^m S^m]\exp[i\frac{1}{M^2}\sum_k J^k S^k]\approx
\nonumber\\
&&\int dJ^k d S^m \det[\frac{1}{M^9}A^kB^mJ^kS^m]\exp[i\sum_k \frac{1}{M^2}J^k S^k]\approx
\nonumber\\
&&\int d J^k d S^m d \bar{z}_i d z_i\exp[i\bar{z}[\frac{1}{M^6}A^kB^mJ^kS^m]z+i\sum_k \frac{1}{M^2}J^k S^k]=
\nonumber\\
&&\int d S^m d \bar{z}_i d z_i\prod_k\delta(\frac{1}{M^2}S^k+\bar{z}\frac{1}{M^6}A^kB^mS^mz)\approx
\nonumber\\
&&\int d \bar{z}_i d z_i\prod_k\frac{1}{1+\frac{1}{M^4}\bar{z}A^kB^kz}\approx\
\nonumber\\
&& \int d \bar{z}_i d z_i\exp[-\sum_k\ln[1+\frac{1}{M^4}\bar{z}A^kB^kz]]\approx
\nonumber\\
&&\int d \bar{z}_i d z_i\exp[-\sum_k\frac{1}{M^4}\bar{z}A^kB^kz]\approx \det[A^kB^k]
\label{res6645}
\end{eqnarray}
Here we dropped all unimportant constant factors and in the last line we consider only terms in order $\frac{1}{M^4}$ and dropped the higher order ones in the fermion fields. This is the only approximation involved in Eq. (\ref{res6645}).

Now consider again $Z_x$ and this time we integrate first over $J_k$ and $S_k$:
\begin{eqnarray}
&&Z_x= \int d \bar{x}_i d x_i d \bar{y}_j d y _j d J_k d S_m\exp[i[\frac{1}{M^5}\bar{x}A^kxJ^k+\frac{1}{M^4}bar{y}B^kyS^k+\sum_{k}\frac{1}{M^2}S^kJ^k]]\approx
\nonumber\\
&&=\int d \bar{x}_i d x_i d \bar{y}_j d y _j d J_k \prod_p\delta(J^p+\frac{1}{M^2}\bar{y}B^py)\exp[i\frac{1}{M^5}\bar{x}A^kxJ^k]\approx
\nonumber\\
&&\int d \bar{x}_i d x_i d \bar{y}_j d y _j \exp[-i\frac{1}{M^7}\bar{x}A^kx\bar{y}B^ky]\approx\det[A^mB^m].
\label{rez55456}
\end{eqnarray}
The result in the last line of Eq. (\ref{rez55456}) will be of most importance in what follows.

There is an important extension to the results in Eqs. (\ref{res6645}) and (\ref{rez55456}) which we will  use here but state without proof because the proof is just a simple generalization of the arguments above:
\begin{eqnarray}
&&Z_z= \int d \bar{x}_i d x_i d \bar{y}_j d y _j d J_k d S_m\exp[i[\frac{1}{M^5}\bar{x}A^kxJ^k-\frac{1}{M^4}\bar{y}B^kyS^k+\frac{1}{M^2}\sum_{k}S^kJ^k+\frac{1}{M^3}\bar{x}Cx]]=
\nonumber\\
&&=\int d \bar{x}_i d x_i d \bar{y}_j d y _j \exp[i\frac{1}{M^7}[\bar{x}A^kx\bar{y}B^ky]]\approx\det[B^mA^m].
\label{veryim8867}
\end{eqnarray}
Here $C$ is an $n\times n$ matrix but one would obtain the same result if $C$ were a polynomial with terms of the type $(\bar{x}Rx)^k$ with $R$ any $n\times n$ matrix  and k an arbitrary integer. The reason stems from matching the degrees of freedom of Grassmann variables $x$ and $y$.

We thus start from:
\begin{eqnarray}
&&Z_x= \int d \bar{x}_i d x_i d \bar{y}_j d y _j d J_k d S_m\exp[i\int d^4 x[\frac{1}{M}\bar{x}A^kxJ^k-\bar{y}B^kyS^k+\sum_{k}M^2S^kJ^k+T(x,\bar{x})]]=
\nonumber\\
&&=\int d \bar{x}_i d x_i d \bar{y}_j d y _j\exp[i\int d^4 x[\frac{1}{M^3}\bar{x}A^kx\bar{y}B^ky+T(x,\bar{x})]]\approx \det[B^kA^k],
\label{intyyt2u644}
\end{eqnarray}
and consider the set $x_i$ as being $N\frac{N_f}{2}$ fermions $\Psi_{if}$ , where $N$ is the number of colors and $\frac{N_f}{2}$ is the number of flavors. Similarly the set $y_i$ is a similar set of $N\frac{N_f}{2}$
fermions $\chi_{if}$. In total we consider to have $N_f=6$ flavors of fermions  corresponding to the $6$ flavors of quarks of the standard. We further define the matrices $A^k$, $B^k$ as:
\begin{eqnarray}
&&A^k=[i\partial_{\mu}t^a-mt^a\gamma_{\mu}+g'A^a_{\mu}]
\nonumber\\
&&B^k=\gamma^{\mu}t^a\\
\nonumber\\
&&T=\frac{1}{m_0^2}(\sum_f\bar{\Psi}_f\Psi_f-v^3)^2,
\label{rez66578}
\end{eqnarray}
where $v$ is a constant with mass dimension $1$ and $m_0$ is an arbitrary scale.
Then:
\begin{eqnarray}
&&B^kA^k==[i\gamma^{\mu}\partial_{\mu}t^at^a-4mt^at^a+g'\gamma^{\mu}t^aA^a_{\mu}]=
\nonumber\\
&&[i\gamma^{\mu}\partial_{\mu}\frac{N^2-1}{2N}-4m\frac{N^2-1}{2N}+g'\gamma^{\mu}t^aA^a_{\mu}]=
\nonumber\\
&&\frac{N^2-1}{2N}[i\gamma^{\mu}\partial_{\mu}-4m+g\gamma^{\mu}t^aA^a_{\mu}]
\label{res554664}
\end{eqnarray}
where we redefined $g'=\frac{N^2-1}{2N}g$ and $4m=m_f$. We observe that the operator $B^kA^k$ is the operator that appears in the standard model between two fermion states. Note that we can include in each $A^k$ and $B^k$ a diagonal flavor matrix with the same final result.

Then the counterpart of the Eq. (\ref{intyyt2u644}) in terms of the above definition will be:
\begin{eqnarray}
&&Z_0=\int d A^a_{\mu}(x)\prod_f\det\left[i\gamma^{\mu}\partial_{\mu}-m_f+g\gamma^{\mu}A^a_{\mu}\right]\exp\left[i[\int d^4x {\cal L}_2+\int d^4 x T]\right]=
\nonumber\\
&&=\int d A^a_{\mu}\prod_{f=1}^3 d \bar{\Psi}_f d \Psi d \bar{\chi}_f d \chi_f\exp\left[i\int d^4x {\cal L}_2\right]\times
\nonumber\\
&&\exp\left[i[\frac{1}{M^3}\int d^4 x\sum_{f=1}^3[\bar{\Psi}_f(it^a\partial_{\mu}-mt^a\gamma_{\mu}+g'A^a_{\mu})\Psi_f]\sum_{f'=1}^3[\bar\chi_{f'}\gamma^{\mu}t^a\chi_{f'}]+\int d^4x
T]\right].
\label{fulllagr}
\end{eqnarray}
Here $M$ is an arbitrary scale that reestablishes the correct dimensionality.

We denote:
\begin{eqnarray}
Z_1=\int \prod_{f=1}^3 d \bar{\Psi}_f d \Psi_f  d \bar{\chi}_f d \chi_f
\exp\left[i[\frac{1}{M^3}\int d^4 x\sum_{f=1}^3[\bar{\Psi}_f(it^a\partial_{\mu}-mt^a\gamma_{\mu}+g'A^a_{\mu})\Psi_f]\sum_{f'=1}^3[\bar\chi_{f'}\gamma^{\mu}t^a\chi_{f'}]+\int d^4 x T]\right],
\label{part45546}
\end{eqnarray}
and work only with it.

First we will make the change of variable $J^a_{\mu f}=\frac{1}{M^2}\bar{\chi}_f\gamma^{\mu}t^a\chi_f$ (see section II for details). Note that this implies a subsequent gauge condition for $J^a_{\mu f}$ which we shall discuss later. The partition function $Z_1$ will become:
\begin{eqnarray}
Z_1=\int \prod_{f=1}^3 d \bar{\Psi}_f d \Psi_f d J^a_{\mu f} \det[J^a_{\mu f}\gamma^{\mu}t^a]
\exp\left[ i [\frac{1}{M}\int d^4 x\sum_{f=1}^3\bar{\Psi}_f(it^a\partial_{\mu}-mt^a\gamma_{\mu}+g'A^a_{\mu})\Psi_f \sum_{f=1}^3J^{a\mu f}+\int d^4x T]\right].
\label{part45546}
\end{eqnarray}
Noting that in the change of variable we can use $\det[\sum_fJ^a_{\mu f}\gamma^{\mu}t^a]^3$ instead of $\prod_f\det[J^a_{\mu f}\gamma^{\mu}t^a]^3$ (see Eq. (\ref{res53453}) we can further write:
\begin{eqnarray}
Z_1=\int \prod_{f=1}^3 d \bar{\Psi}_f d \Psi_f d \Psi_f d J^a_{\mu f} \det[\sum_f J^a_{\mu f}\gamma^{\mu}t^a]^3
\exp\left[i[\frac{1}{M}\int d^4 x\sum_{f=1}^3[\bar{\Psi}_f(it^a\partial_{\mu}-mt^a\gamma_{\mu}+g'A^a_{\mu})\Psi_f] \sum_{f=1}^3J^{a\mu f}+\int d^4x T]\right].
\label{pareet245546}
\end{eqnarray}
We can further make a change of variables $Y^a_{\mu }=\sum_{f=1}^3 J^a_{\mu f}$, $Z^a_{\mu }=J^a_{\mu 2}$, $U^a_{\mu 3}=J^a_{\mu 3}$ and drop the unwanted integrals over $Z^a_{\mu }$ and $U^a_{\mu }$ as they would not contribute to any process. This yields:
\begin{eqnarray}
&&Z_1\approx \int dY^a_{\mu} \int \prod_{f=1}^3 d \bar{\Psi}_f d\Psi_f \det[ Y^a_{\mu }\gamma^{\mu}t^a]^3\exp\left[i[\frac{1}{M}\int d^4 x\sum_{f=1}^3[\bar{\Psi}_f(it^a\partial_{\mu}-mt^a\gamma_{\mu}+g'A^a_{\mu})\Psi_f] Y^{a\mu}+\int d^4x T]\right]=
\nonumber\\
&&\int dY^a_{\mu} \int \prod_{f=1}^3 d \bar{\Psi}_f d\Psi_f d\bar{\xi}_f d\xi_f\times
\nonumber\\
&&\exp\left[i[\frac{1}{M}\int d^4 x\sum_{f=1}^3[\bar{\Psi}_f(it^a\partial_{\mu}-mt^a\gamma_{\mu}+g'A^a_{\mu})\Psi_f] Y^{a\mu}+a\sum_{f=1}^3\bar{\xi}_f\gamma_{\mu}t^a\xi_fY^{a\mu}+\int d^4 x T]\right].
\label{res554647}
\end{eqnarray}
 Here we introduced another set $\bar{\xi}_f$ and $\xi_f$ of $3$ fermions to account for the determinant in the first line of Eq. (\ref{res554647}). Here $a$ is an arbitrary dimensionless coupling constant

As we mentioned previously in order to be able to make a change of variable from the fermions $\chi_f$ to the currents $J^a_{\mu f}$ one would need something similar a to gauge condition to cut off the number of degrees of freedom from $32$ to $24$. We shall consider this constraint as $\partial^{\mu}J^a_{\mu}=0$. Upon the change of variables to $Y^a_{\mu}$, $Z^a_{\mu}$ and $U^a_{\mu}$ this constraint will become $\partial^{\mu}Y^a_{\mu}=0$ for the only variable of interest. We shall introduce this in the partition function $Z_1$ as:
\begin{eqnarray}
\delta(\partial^{\mu}Y^a_{\mu})=\int d S^a\exp[i\int d^4x M_1S^a\partial^{\mu}Y^a_{\mu}]=\int d S^a \exp[-i\int d^4xM_1\partial^{\mu}S^aJ^a_{\mu}],
\label{res24434345}
\end{eqnarray}
where $M_1$ is an arbitrary constant with mass dimension $1$.
With the addition of the gauge condition the partition function in Eq. (\ref{res554647}) will become:
\begin{eqnarray}
&&Z_1\approx \int dY^a_{\mu}d S^a \int \prod_{f=1}^3 d \bar{\Psi}_f d\Psi_f \det[ Y^a_{\mu }\gamma^{\mu}t^a]^3\exp\left[i[\frac{1}{M}\int d^4 x\sum_{f=1}^3[\bar{\Psi}_f(it^a\partial_{\mu}-mt^a\gamma_{\mu}+g'A^a_{\mu})\Psi_f] Y^{a\mu}+\int d^4 x
T]\right]=
\nonumber\\
&&\int dY^a_{\mu} \int \prod_{f=1}^3 d \bar{\Psi}_f d\Psi_f d\bar{\xi}_f d\xi_f\times
\nonumber\\
&&\exp\left[i\int d^4 x[\frac{1}{M}\sum_{f=1}^3[\bar{\Psi}_f(it^a\partial_{\mu}-mt^a\gamma_{\mu}+g'A^a_{\mu})\Psi_f] Y^{a\mu}+a\sum_{f=1}^3\bar{\xi}_f\gamma_{\mu}t^a\xi_fY^{a\mu}-
 M_1\partial_{\mu}S^aY^{a\mu}+ T]\right].
\label{re221s554647}
\end{eqnarray}
We shall integrate $Y^{a\mu}$ by observing that it couples only linearly that the Lagrangian is hermitian and thus it leads to a product of delta functions. This yields:
\begin{eqnarray}
&&Z_1=  \int dS^a\prod_{f=1}^3 d \bar{\Psi}_f d\Psi_f d\bar{\xi}_f d\xi_f \times
 \nonumber\\
&&\delta(g'A^a_{\mu}\sum_f\bar{\Psi}_f\Psi_f+\sum_{f=1}^3[\bar{\Psi}_f(it^a\partial_{\mu}-mt^a\gamma_{\mu})\Psi_f]+Ma\sum_{f=1}^3\bar{\xi}_f\gamma_{\mu}t^a\xi_f-M_1M\partial_{\mu}S^a)\times
\nonumber\\
&&\exp[i\int d^4x T]
\label{res443566}
\end{eqnarray}
We shall regard the delta function in the space of variables $A^a_{\mu}$.

Before going further we need to make an important ammendment. When we introduce the change of variable from $\chi_f$, $\bar{\chi}_f$ to $J^a_{\mu f}$ we replaced (we consider here for simplicity only one fermion species):
\begin{eqnarray}
\int  d \bar{\chi}_f d \chi_f\approx \int d J^a_{\mu f}\det[\gamma^{\mu}t^aJ^a_{\mu f}]
\label{res55454}
\end{eqnarray}
The reason stems form the fact that when we transform from one set of variables one gets always products of $24$ different fermion components. However one can extend the above transformation to include some function in the determinant. Consider that we have:
\begin{eqnarray}
\int d J^a_{\mu}\det[i\gamma^{\mu}\partial_{\mu}-m+kt^a\gamma^{\mu}J^a_{\mu}]
\label{res77565}
\end{eqnarray}
and we apply the inverse transformation to fermion variables $\chi_f$:
\begin{eqnarray}
&&\int d J^a_{\mu}\det[i\gamma^{\mu}\partial_{\mu}-m+kt^a\gamma^{\mu}J^a_{\mu f}]=
\nonumber\\
&&\int   d \bar{\chi}_f d \chi_f |\frac{d J^a_{\mu}}{d\chi_f}|^{-1}\det[i\gamma^{\mu}\partial_{\mu}-m+kt^a\gamma^{\mu}\bar{\chi}_f \gamma_{\mu}t^a\chi_f]
\label{tre34}
\end{eqnarray}

Moreover we also need to consider a function (for one flavor) to be integrated which is of the type $\det[X^a_{\mu}J^a_{\mu}]$. Assume we express this in terms of fermion components. Those variables that repeat themselves will get canceled. In the end $\det[X^a_{\mu}J^a_{\mu}]=f\bar{\chi}_1\bar{\chi}_2..\chi_1\chi_2$ where f is an arbitrary function and the product is over all $24$ fermion components for one flavor (note that the product might contain derivative which we omit for simplicity). Then Eq. (\ref{tre34}) will become:
\begin{eqnarray}
&&\int d J^a_{\mu}\det[i\gamma^{\mu}\partial_{\mu}-m+kt^a\gamma^{\mu}J^a_{\mu f}]\det[X^a_{\mu}J^a_{\mu}]=
\nonumber\\
&&\int d \bar{\chi} d\chi\frac{1}{\bar{\chi}_1\bar{\chi}_2...\chi_1\chi_2..}[a_0+a_1\bar{\chi}_i\chi_j+...a_{12}\bar{\chi}_1\bar{\chi}_2...\chi_1\chi_2..]\times[\bar{\chi}_1\bar{\chi}_2...\chi_1\chi_2..]=
\nonumber\\
&&\int d \bar{\chi} d\chi[a_0+a_1\bar{\chi}_i\chi_j+...a_{12}[\bar{\chi}_1\bar{\chi}_2...\chi_1\chi_2..]]=\int d \bar{\chi} d\chi[a_{12}\bar{\chi}_1\bar{\chi}_2...\chi_1\chi_2..]
\label{re435536}
\end{eqnarray}
But the end result corresponds to that part of $\det[i\gamma^{\mu}\partial_{\mu}-m+kt^a\gamma^{\mu}J^a_{\mu f}]$ that contains only $J^a_{\mu}$ so the addition we make to $\det[t^a\gamma^{\mu}J^a_{\mu}]$ is irrelevant.  The main point of the above discussion is that even when we integrate over $J^a_{\mu}$ the intrinsic fermion nature of the variables should be considered. For most of the purposes we shall consider $J^a_{\mu}$ however as a regular commuting variable. However our discussion has a counterpart by discussing purely the $J^a_{\mu}$'s. This being settled one can add in  Eq. (\ref{res443566}) a kinetic term for the fermions $\bar{\xi}_f$, $\xi_f$ as in:
\begin{eqnarray}
&&Z_1=\int  \int \prod_{f=1}^3 d \bar{\Psi}_f d\Psi_f d\bar{\xi}_f d\xi_f dS^a\times
\nonumber\\
&&\prod
\delta(g'A^a_{\mu}\sum_f\bar{\Psi}_f\Psi_f+\sum_{f=1}^3[\bar{\Psi}_f(it^a\partial_{\mu}-mt^a\gamma_{\mu}\Psi_f]+Ma\sum_{f=1}^3\bar{\xi}_f\gamma_{\mu}t^a\xi_f-M_1\partial_{\mu}S^a))\times
\nonumber\\
&&\exp[i[\int d^4 x [\sum_f [\bar{\xi}_f(i\gamma^{\mu}\partial_{\mu}-M_f)\xi_f]+ \int d^4 x T]].
\label{r2es443566}
\end{eqnarray}
The expression in Eq. (\ref{r2es443566}) is the final partition function we will work with.

It appears that the delta function in Eq. (\ref{r2es443566}) is badly defined. In order to fix this we go back to the full Lagrangian of interest (including the pure gluon one) which is:
\begin{eqnarray}
&&{\cal L}=\sum_f [\bar{\xi}_f(i\gamma^{\mu}\partial_{\mu}-M_f)\xi_f]+ T+
\nonumber\\
&&J^a_{\mu}[g'A^a_{\mu}\sum_f\bar{\Psi}_f\Psi_f+\sum_{f=1}^3[\bar{\Psi}_f(it^a\partial_{\mu}-mt^a\gamma_{\mu}\Psi_f]+Ma\sum_{f=1}^3\bar{\xi}_f\gamma_{\mu}t^a\xi_f-M_1M\partial_{\mu}S^a)]+
{\cal L}_2(A^a_{\mu},g)
\label{fulllagr4554}
\end{eqnarray}
We first recall that $g'=\frac{N^2-1}{2N}g$ and make the change of variables $gA^a_{\mu}\Rightarrow A^a_{\mu}$. Note that we can do this at any time without affecting in any way the dynamics.
This yields (we use for simplicity the same notation):
\begin{eqnarray}
&&{\cal L}=\sum_f [\bar{\xi}_f(i\gamma^{\mu}\partial_{\mu}-M_f)\xi_f+ T+
\nonumber\\
&&J^a_{\mu}[\frac{N^2-1}{2N}A^a_{\mu}\sum_f\bar{\Psi}_f\Psi_f+\sum_{f=1}^3[\bar{\Psi}_f(it^a\partial_{\mu}-mt^a\gamma_{\mu}\Psi_f]+Ma\sum_{f=1}^3\bar{\xi}_f\gamma_{\mu}t^a\xi_f-M_1M\partial_{\mu}S^a)]+
\frac{1}{g^2}{\cal L}_2'(A^a_{\mu}),
\label{fu22lllagr4554}
\end{eqnarray}
where ${\cal L}_2'$ does not contain any more any trace of $g$. We are interested in the regime where $g$ is large so we can consider $\frac{1}{g^2}$ a small parameter.
Next we take into account that one can solve from the delta function $A^a_{\mu}$ as  function of the other variables. With this substitution the Lagrangian becomes:
\begin{eqnarray}
&&{\cal L}=\sum_f [\bar{\xi}_f(i\gamma^{\mu}\partial_{\mu}-M_f)\xi_f]+ T+
\frac{1}{g^2}{\cal L}_2'(\bar{\xi}_f,\xi_f,\bar{\Psi}_f, \Psi_f,S^a)
\nonumber\\
&&{\cal L}_0=\sum_f [\bar{\xi}_f(i\gamma^{\mu}\partial_{\mu}-M_f)\xi_f]+ T,
\label{fu3322lllagr4554}
\end{eqnarray}
Assume we consider separately shifts in the variables $\xi_f$, $\Psi_f$ and their subsequent conjugates:
\begin{eqnarray}
&&\xi_f=\xi_f+\frac{1}{g^2}\xi_f'
\nonumber\\
&&\bar{\xi}_f=\xi_f+\frac{1}{g^2}\bar{\xi}_f'
\nonumber\\
&&\Psi_f=\xi_f+\frac{1}{g^2}\Psi_f'
\nonumber\\
&&\bar{\Psi}_f=\xi_f+\frac{1}{g^2}\bar{\Psi}_f'.
\label{sh7756}
\end{eqnarray}
Then one can still obtain the interaction Lagrangian of order $\frac{1}{g^2}$ provided that the fields $\Psi_f$ and $\xi_f$ satisfy the equation of motion for the free Lagrangian ${\cal L}_0$.  Recalling the definition of $T$ from Eq. (\ref{rez66578}) we get:
\begin{eqnarray}
\frac{\partial T}{\partial \Psi}=\frac{2}{m_0^3}\Psi(\sum_f\bar{\Psi}_f\Psi_f-v^3)=0
\label{yt665}
\end{eqnarray}
whose only non trivial solution is $\sum_f\bar{\Psi}_f\Psi_f=v^3$. Thus the unwanted denominator in the $\delta$ function in Eq. (\ref{r2es443566}) is conveniently fixed.

We redefine $\frac{1}{v^3}M=y\frac{1}{v^2}$ and $\frac{1}{v^3}MM_1=z\frac{1}{v}$ where $y$ and $z$ are two adimensional coefficients.

From Eqs. (\ref{r2es443566}) and (\ref{fu3322lllagr4554}) we obtain  the effective fermion Lagrangian in terms of the two sets of fermions $\Psi_f$ (the light quarks) and $\xi_f$ (the heavy quarks):
\begin{eqnarray}
{\cal L} =\sum_f [\bar{\xi}_f(i\gamma^{\mu}\partial_{\mu}-M_f)\xi_f]+\frac{1}{m_0^2}(\sum_f\bar{\Psi}_f\Psi_f-v^3)^2+
\frac{1}{g^2}{\cal L}_2'(A^a_{\mu}(\Psi_f,\xi_f,S^b)),
\label{fin677}
\end{eqnarray}
where,
\begin{eqnarray}
A^a_{\mu}(\Psi_f,\xi_f,S^b)=-\frac{2N}{N^2-1}[\frac{1}{v^3}\sum_{f=1}^3\bar{\Psi}_f(it^a\partial_{\mu}-mt^a\gamma_{\mu})\Psi_f+
y\frac{1}{v^2}\sum_{f=1}^3\bar{\xi}_f\gamma_{\mu}t^a\xi_f-z\frac{1}{v}\partial_{\mu}S^a].
\label{expt56657}
\end{eqnarray}
We can further refine Eq. (\ref{expt56657}) by taking $A^a_{\mu}$ equal to  the real part of the right hand side which leads to the additional constraint on the field $\Psi_f$:
\begin{eqnarray}
\partial_{\mu}(\sum_{f=1}^3 \bar{\Psi}_ft^a\Psi_f)=0.
\label{res1112}
\end{eqnarray}

In Eq. (\ref{fin677}) the term ${\cal L}_2'$ is the pure gluon Lagrangian independent of the gauge coupling constant:
\begin{eqnarray}
{\cal L}_2'=-\frac{1}{4}F^{a\mu\nu}F^a_{\mu\nu}.
\label{lag5665}
\end{eqnarray}
Note   that in Eq. (\ref{expt56657})  the presence of the scalars $S^a$ ($a=1...8$) is purely optional as there are  many ways in which one can implement this constraint. Also if $M_f$ are considered large one can integrate out the $\xi_f$ fermions this leading to an equivalent Lagrangian expressed only in terms of the light degrees of freedom $\Psi_f$.

\section{Discussion}

When one constructs an effective theory for the low energy degrees of freedom of QCD, be these fermions or  hadrons, one uses often as a guiding principle the approximate symmetries  already established experimentally such as the chiral $SU(3)_L\times SU(3)_R$ symmetry or the $U(1)_A$ axial one. In the present work we considered a different approach; thus instead of dwelling on  symmetries we focused on the intrinsic dynamics encapsulated in the partition function of the QCD Lagrangian. Using alternative sets of variables we were able to reproduce the exact  partition function obtained after integrating out the quark degrees of freedom. It turns out that a  partition function depending on a set of fermion variables can be described alternatively only by new sets of fermion variables, consider them copies of the first ones, that however have a completely different Lagrangian.

The final result (see Eq. (\ref{fin677})) is an effective Lagrangian  with unusual terms depending on two groups of fermions (where $\Psi_f$ are the three light quarks and $\xi_f$ are the three heavy ones) and with couplings which go up to $8$ fermion interaction terms.
This Lagrangian gives an adequate partition function corresponding to that of the original QCD Lagrangian for the case when the coupling constant $g$ is large.  Moreover it has an intrinsic $SU(3)_L\times SU(3)_R$ chiral symmetry in the two sets of fermion sectors for the case when the quark masses $m_f$ and $M_f$ are set to zero and also an $SU(3)_V$ symmetry for the case when the two sets of masses are equal within one set.

The purely quark Lagrangain we obtained can be further processed by using QCD rules to extract the bound states of mesons or baryons and their interactions. However the overall method goes far beyond this. One could at any intermediate stage introduce directly instead of currents the meson states. In this case one must take into account that initially there are $72$ degrees of freedom for the light quarks corresponding to a product of $3$ colors, $3$ flavors and $8$ space time coordinates of an off-shell fermion. This means that the partition function must accommodate only that number of mesons, be they scalars, pseudoscalar, vectors etc. whose total number of degrees of freedom sum up $72$.  All alternatives that conveniently express the initial Lagrangian should be considered and all Lagrangians that can be constructed in this way are possible outcomes. This is also true for the Lagrangian in Eq. (\ref{fin677}) as it is only one of the many possible choices compatible with the initial QCD partition function.

In this paper we propose a new theoretical effective fermion Lagrangian obtained from QCD not by integrating out the gluons but by eliminating these as a result of mocking up the exact partition function of QCD. We shall leave the discussion of phenomenological implications of our model for further work.

\section*{Acknowledgments} \vskip -.5cm

The work of R. J. was supported by a grant of the Ministry of National Education, CNCS-UEFISCDI, project number PN-II-ID-PCE-2012-4-0078.

\end{document}